# Wind Tunnel Study of the Forces Due to Drafting in Dolphin Mother-Calf pairs

D. Weihs and M. Ringel

Faculty of Aerospace Engineering and Autonomous Systems Program

Technion, Haifa 32000 Israel

## Introduction

Understanding the hydrodynamic interactions between mother cetaceans and their calves ( Figure 1) has been the motivation for studies over several decades. This was initiated by the observation that calves that are usually assisted by drafting with their mothers during fast swimming, get separated from their mothers during escape from tuna fishing nets ( Ballance et al. 2021, Edwards,2006, Noren ,2013). This is thought to contribute to the losses to dolphin populations (Gerrodette and Forcada 2005) , even after the  net escape procedure was mandated  over 20 years earlier.

While theoretical analysis defined the problem, experimental data has been extremely difficult to collect in the wild. Experimental evidence was obtained from aerial photography and other observations, but quantitative controlled data is difficult to measure. Some visual data can be obtained from captive dolphins but the coherent swimming distances are not sufficient.

In a previous theoretical study by the senior author (Weihs, 2004) on dolphin mother-calf interactions, , the presence of a neighboring body was shown to result in side-forces of the order of the drag force on each body (known as Bernoulli attraction), and by suitable spacing in the longitudinal direction, the rear body was shown to experience much lower drag, even obtaining complete cancellation of drag for calves next to adult dolphins, allowing them to coast. In parallel, wind tunnel studies of similar elongated bodies for a different purpose gave excellent agreement with the theoretical results ( Weihs et al. 2006).

These results compared well with observations of Eastern Spinner dolphin mother-calf paired swimming even though the models used, in theory and experiments were geometrically simplified to equal-sized elongated bodies of revolution.

Following the renewed interest in this problem (Pearson et al., 2023] by the Inter-American Tropical Tuna Commission, we saw the opportunity to use experiments in the wind tunnel to obtain controlled data for the forces involved when dolphin mother-calf pairs swim in close proximity.

Applying the analysis in (Weihs, 2004) which showed good agreement with observations, we designed and built three models of spheroidal shape and aspect ratio 6, based on data for adult and neonate Spotted dolphins (*Stenella attenuata*) . The models have different nominal lengths- 540mm, 270 mm and 135 mm ( Figure 2a). The middle sized model is identical with the models from previous work mentioned above (Weihs et al. 2006) , so that calibration could be based on previous experimental data, saving valuable tunnel run-time. These models were run in the Technion Wind Tunnel Laboratory. While wind tunnel results do not fully reproduce motion in the sea, as surface effects, and added mass contributions are not represented, they can verify the predictions of the basic theoretical model, which did not include surface effects. As such the wind tunnel results spotlight the interaction effects between the bodies. A series of runs was performed, and analyzed, in which two models of differing sizes were run at several lateral and longitudinal relative placements, to verify the predictions of the dependence of size and relative location on the forces on the dolphin mother and calf.

Theoretical Background

We start by modeling the dolphin shape, when coasting, as having an elongate shape i.e. the length to diameter ratio l/d>>1 where d is the maximum diameter and l the length. As a result, the flow field around such a body can be assumed to vary slowly in the longitudinal direction. Following the analysis in Weihs, (2004), such bodies are defined as slender, allowing a simplified set of equations. This assumption can be shown to be increasingly accurate for slenderness ratios of over 4. In our case, the slenderness ratio is 6, i.e. well within the range of validity of the model. The speeds at which the tests took place are low enough so that compressibility effects can be neglected, so that the results in air represent results in water, albeit away from the surface. Neonates of several dolphin species are similar in shape to adults, and roughly half the length of the fully grown animals. To simplify matters further we look at spheroids, i.e. bodies, whose cross sectional area is circular.

The mathematical steps required to get the effects of proximity between mother and calf appear in Weihs (2004). They will not be repeated here, except to show the geometry and the final equations used to calculate the interaction.

The model assumes potential (inviscid) flow, so that a single closed body such as our ellipsoid moving at constant speed will experience zero net force (the D'Alembert paradox). However, when the two bodies are in proximity, the flow asymmetries caused by the presence of the neighboring body results in net forces on each of them, even when viscosity is not taken into account. In our experimental data, These effects are separated , as will be shown later.

The coordinate system used can be defined by (Fig 3, adapted from Weihs, 2004) x is the streamwise direction, y the lateral and z the vertical ( which is not

$$x_1 = x_2 - \xi \tag{1}$$
$$y_1 = y_2 - \eta$$
$$z_0 = z_1 = z_2 = 0$$

Where $\xi$ is the stagger between the bodies and $\eta$ the lateral distance between axes of symmetry of both bodies;. The undisturbed flow speed is U and S are the maximal cross sections of the bodies. Body 1 ( subscript 1) represents the calf, and 2 the mother.

These forces are

$$X = F_x = \frac{\rho U^2}{4\pi} \int_{L_1} S_1'(x_1) \int_{L_2} \frac{S_2'(x_2)(x_2 - x_1 - \xi) dx_2}{[(x_2 - x_1 - \xi)^2 + \eta^2]^{\frac{3}{2}}} dx_1 \qquad (2)$$

and

$$Y = F_y = \frac{\rho U^2 \eta}{2\pi} \int_{L_1} S_1'(x_1) \int_{L_2} \frac{S_2'(x_2) dx_2}{[(x_2 - x_1 - \xi)^2 + \eta^2]^{\frac{3}{2}}} dx_1 \qquad (3)$$

for the longitudinal, and lateral forces on the forward body due to the presence of the other.

Next we insert the specific body shape.

$$S_i(x_i) = D_i(1 - 4(x_i/L_i)^2) \qquad (4)$$

where $D_i$ ( i=1,2,3) is the maximum girth ( diameter) of each at the equator. Figure 4 (adapted from Weihs,2004) shows how the forces change with relative longitudinal (fore-aft) stagger of the two bodies. The curves are for the case where the mother is twice as long as the calf, as mentioned above, i.e. $L_2/L_1=2$.

Oblate spheroids (ellipsoids) of 6:1 slenderness ratio and three sizes, one with nominal length 540 mm, (designated B), one of 270mm ( designated M), and the third of 135 mm (designated S) were manufactured of Aluminum Al 2024. These are held in place in the tunnel by downstream stings to minimize fluid dynamic effects on the model ,with sting holder sections made of maraging steel 300. The models appear in Figure 1b. The rear of each spheroid was hollowed out and a sting holder of diameter 16mm, for the larger models, and 10mm for the smaller model, installed within a hollow of 24mm and 12 mm diameter respectively. Thus, the actual lengths were slightly smaller (see Table 1).

| Model | Maximum diameter (mm) | Nominal length [mm] | Actual lengt |
|---|---|---|---|
| S (Small) | 22.5 | 135 | 124.8 |
| M ( Medium) | 45 | 270 | 249 |
| B (Big) | 90 | 540 | 510 |

Table 1. Wind tunnel model measurements.

Wind Tunnel Experiments.

As mentioned above we use two such combinations B/M and M/S – see Table 1
The models M and S were tested in the Technion wind tunnel No.1. The tunnel has a 600*800 mm cross-section, so that the blockage for both models was less than 0.6% at zero angle of attack (at which the relevant measurements were taken), and less than 2% for angles of up to $30^0$. Forces and moments were measured by means of two 6 degrees of freedom strain gage sting force measurement balances. The first (Balance 6462 ) used for model M had sensitivity of 3.3 µV /gram for side-forces, and 2.9 µV/gram for axial forces (drag or thrust), and the second ( 6DH3) used with model S had sensitivity of 3.6 µV /gram for side-force, and 1.6 µV/gram for axial force . The spheroids were placed in the tunnel with parallel longitudinal axes (Figure 1) and different $\xi$ and $\eta$ (see figure 2 for definitions).

The tunnel was run at a nominal speed of 75 m/s to produce large enough interactive forces. Both models were moved in uniform together through a range of angles of attack. This was required to determine the aerodynamic zero- angle, i.e. the geometric angle at which the drag forces were minimal. The raw data was normalized to counter the differences occurring due to slight differences in actual speeds, temperature and external pressure measured in individual runs.

Following these runs we tested the pair B/M in Technion AE Wind tunnel No. 10, This tunnel has a 1000/1000 mm cross section, required because of the blockage by the big model B. The balances used were 6462, as before for M and 6501 for the large model B. The sensitivity of balance 6501 was 2.8 µV /gram for side-forces and 2.6 µV /gram for axial forces.

Results

Figure 5 shows the change of maximal side force in Newtons, for the two spheroids B and M, situated in parallel, i.e. with the largest diameters on a lateral line normal to the flow, as a function of lateral distance η/L between centerlines. Figure 5a presents the measured data for the two bodies, showing as expected, that the forces are attractive for both bodies. A certain small asymmetry in the forces is seen, of 1-3 centi-Newtons. This force is an effect of the size difference in the models, resulting in asymmetric wakes. It vanishes when equal bodies are tested( Weihs et al., 2006)  The effect increases with decreasing distance between bodies, but is not important as the shapes are a rough approximation of dolphin bodies anyway.

In figure 5b this asymmetry is subtracted, to show the net Bernoulli attractive force between the bodies. These forces, which are compatible with the inviscid theory, decline rapidly with increasing lateral distance between the centerlines of the bodies, as expected from the theory (see figure 5B of Reference 1).

Figures 6a and 6b show the side-force at a fixed lateral distance as function of the longitudinal displacement of the two bodies. Positive ξ indicates that the front of the mother ( body B) is ahead of the calf  body M. Again, a small asymmetry, growing as the lateral distance decreases appeared.

Subtracting the asymmetry, we see (Fig 6b) that the maximum side-force appears at about ξ/L=-0.15,, i.e. when the center of the smaller body S is slightly behind M, instead of at ξ/L=0 as predicted by the inviscid theory. This again is most probably a viscous effect, as the viscous boundary layer grows when moving lengthwise along the body  and may cause the thickest part of the body to appear downstream of the geometrical maximal thickness.

Figure 7  shows the axial force on both bodies, at a given η/L=0.442 and varying longitudinal displacements. As expected from the theory, (fig 3) there is transference of axial force, and under the right displacement, the smaller body ( calf) can gain enough from the other (mother), to nullify the drag and actually obtain a significant forward force (for the calf). Interestingly, placing the calf obliquely forward can reduce the drag on the mother. While this is not an expected behavior, it might occur  if a calf is  playful.  The viscosity appears as a positive addition to the axial force, as seen from the fact that the axial force on both bodies should have been zero- according to the inviscid flow model when ξ=0.

Discussion

The experimental results of figures 5-7 follow the trends predicted by the theoretical model, for both side forces (the Bernoulli attraction) and longitudinal forces. The experimental results show that placing two slender bodies in close proximity results in significant side and longitudinal forces on both. When the calf is placed near the mother it can obtain a significant boost, allowing it to glide effortlessly, at least till the next breach for breathing. This has been observed frequently, The experiments show the magnitude of the viscous corrections to the theoretical predictions, allowing a more accurate use of the theory, by adding the corrections found here , such as the longitudinal backwards displacement of the calf center of mass  for other, similar configurations.

Figures

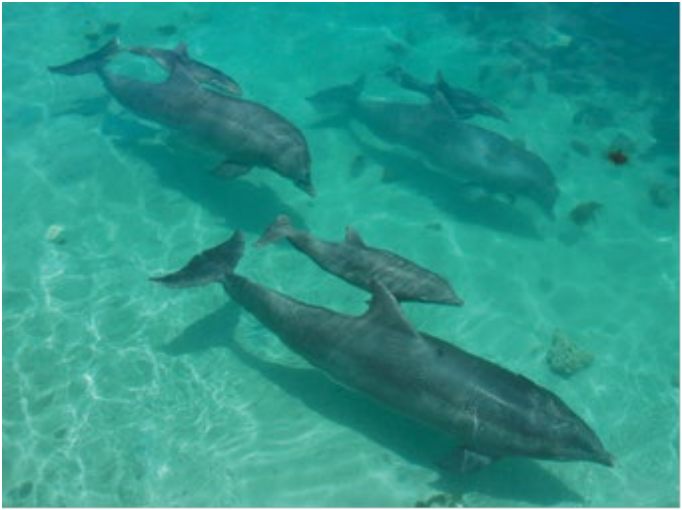

Fig 1  Dolphin Mother-calf pairs

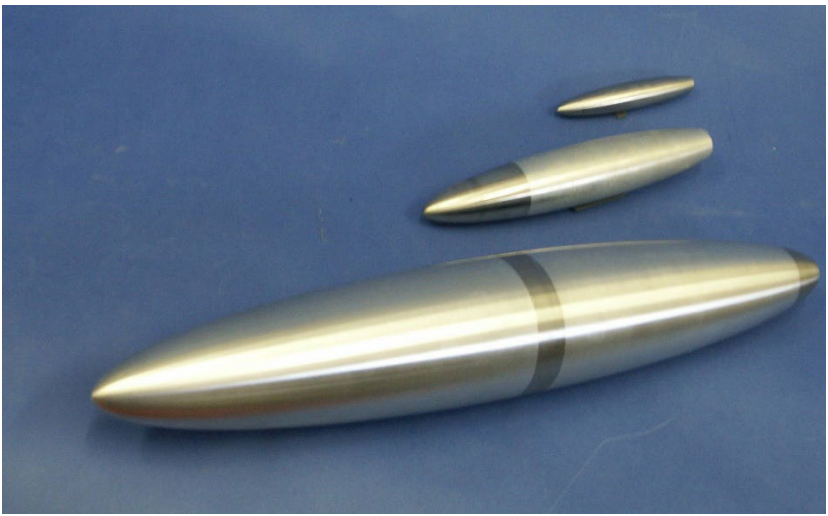

Figure 2a – the wind-tunnel models

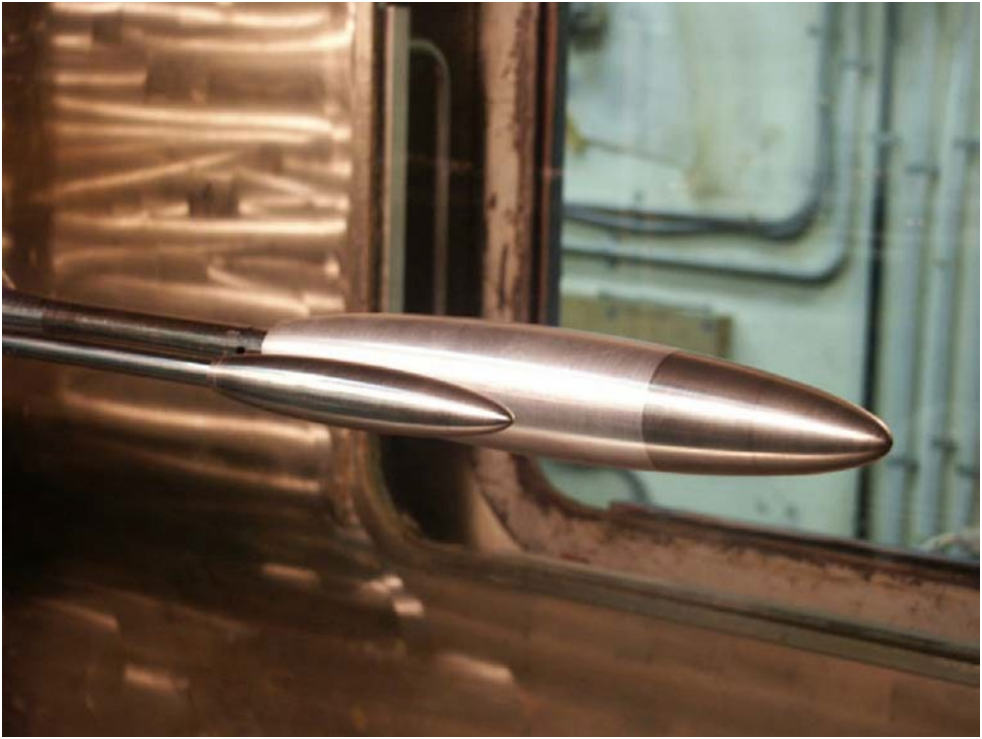

Figure 2b

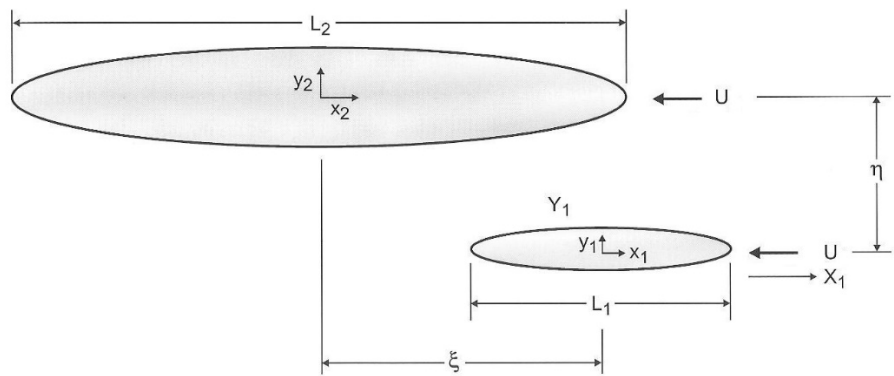

Figure 3 The coordinate system.

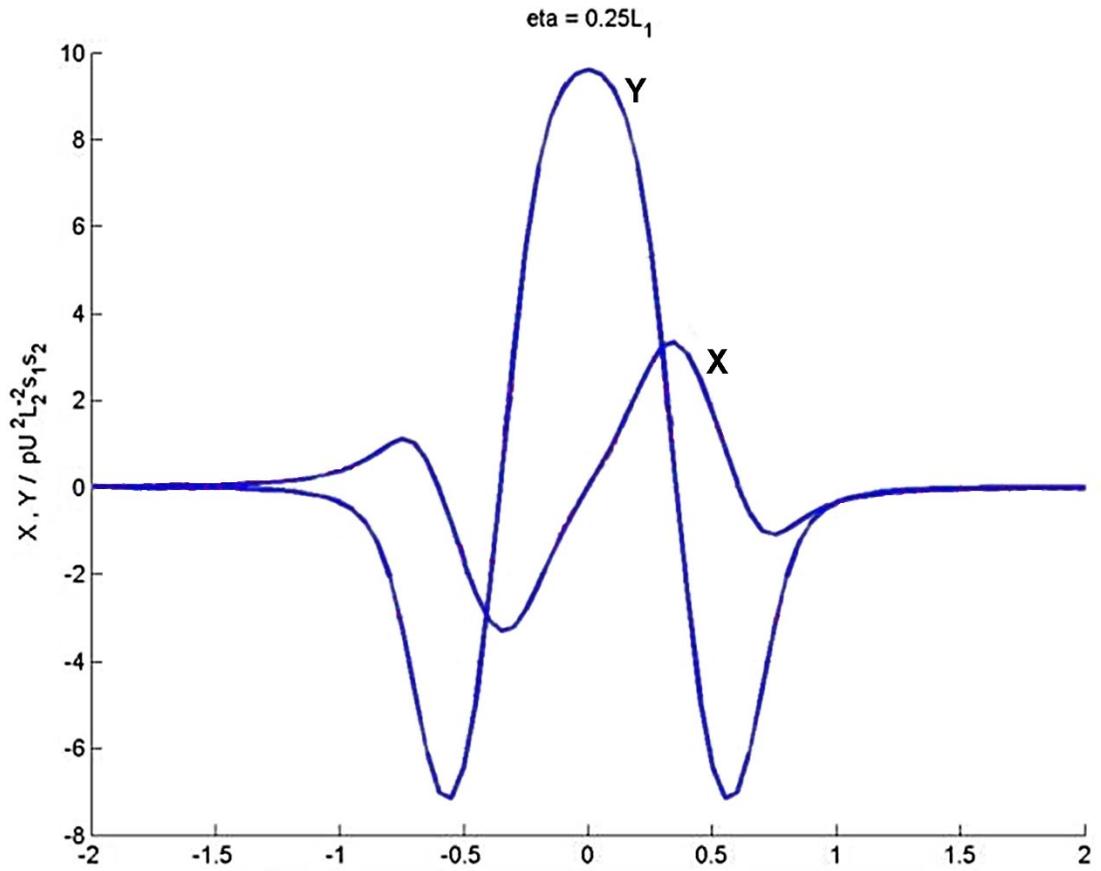

Figure 4 Nondimensional axial X and lateral forces Y on the calf calculated for a lateral spacing of 25% of the smaller body's length (adapted from fig 5 in Weihs, 2004). The horizontal axis is $2\xi/L_1$ so that positive X values indicate that the calf follows the mother.

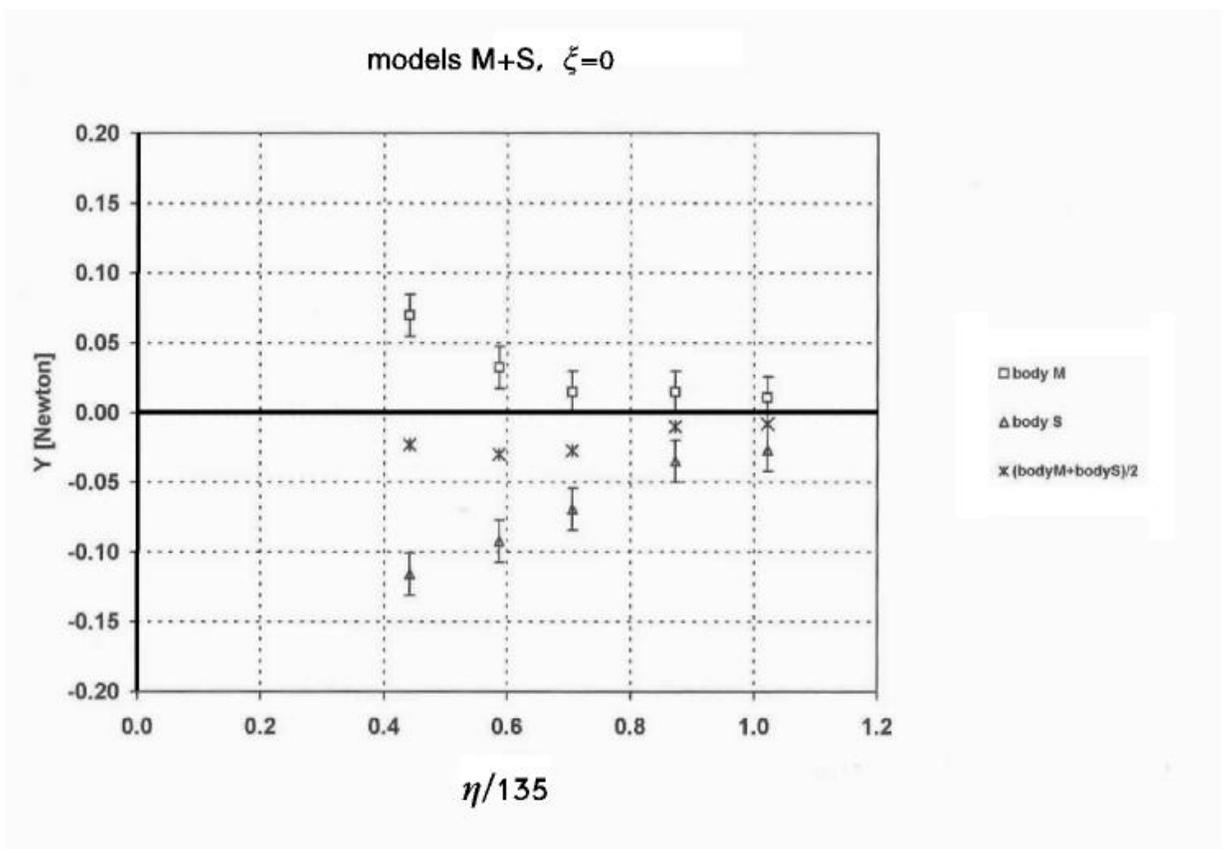

Fig 5a. Side-force on the two bodies, M and S, versus lateral distance. The side-force consists of two contributions, a small asymmetric force, resulting from the size difference of the bodies, marked by x, and a canceling symmetric force.

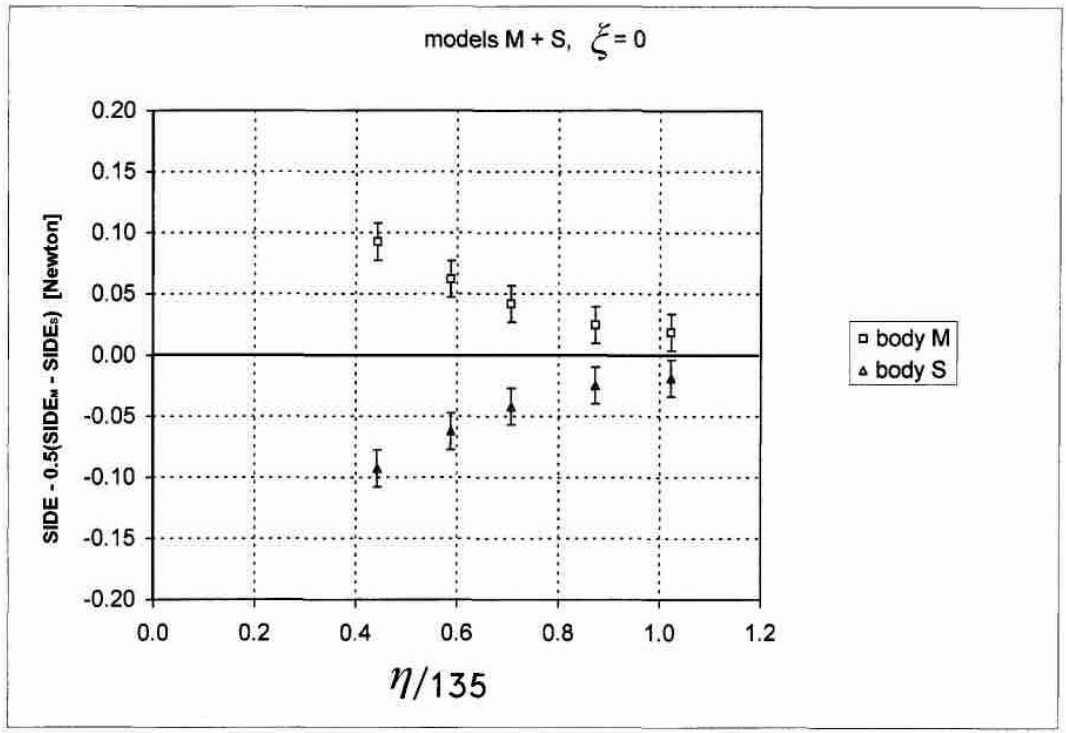

Fig 5b the symmetric side force, obtained by subtracting the asymmetric contribution. This is the potential attractive force, decreasing rapidly with the increase in lateral distance, as predicted by theory

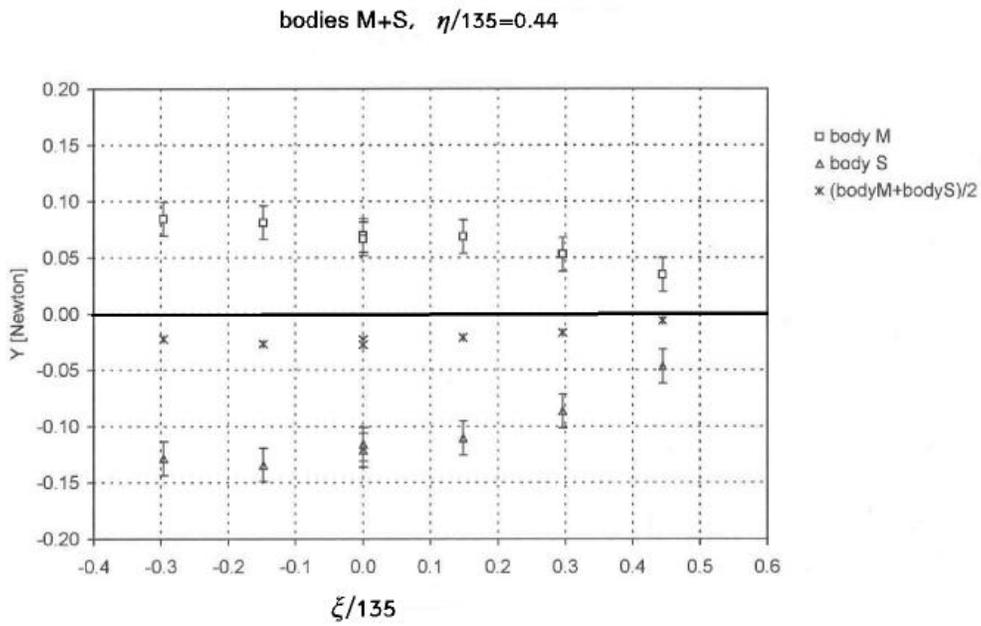

Fig 6a Side-force on the smaller ( S) and larger (M) bodies as a function of normalized longitudinal displacement , and average side-force X.

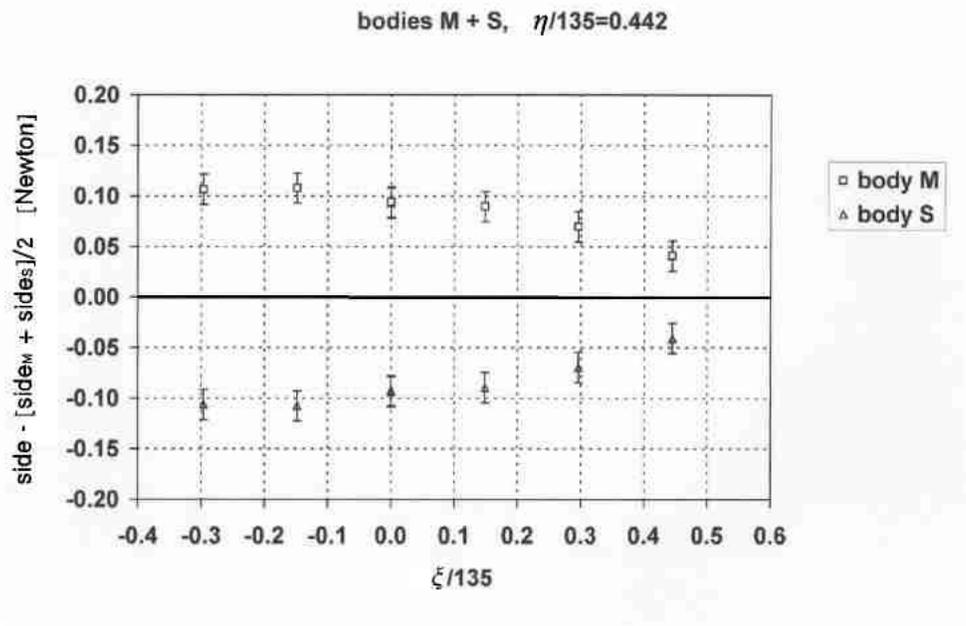

Fig 6b Side-force on the smaller ( S) and larger (M) bodies as a function of normalized longitudinal displacement, with average force subtracted, as I figure 5.

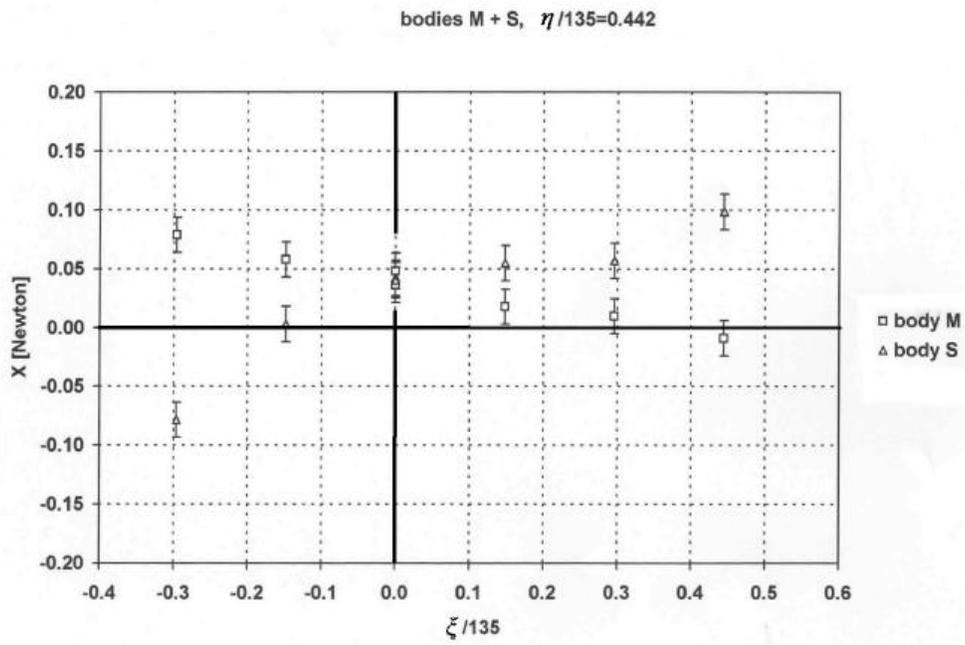

Fig 7 Axial force on the small ( S) and medium (M) bodies, for various longitudinal displacements, and fixed lateral distance.